\documentclass[fleqn,twoside]{article}
\usepackage[headings]{espcrc2}

\readRCS
$Id: espcrc2.tex,v 1.2 2004/02/24 11:22:11 spepping Exp $
\usepackage{graphicx}
\usepackage[figuresright]{rotating}

\newcommand{\AmS}{{\protect\the\textfont2
  A\kern-.1667em\lower.5ex\hbox{M}\kern-.125emS}}

\title{Probing Electroweak Top Quark Couplings at Hadron and Lepton Colliders}

\author{U.~Baur\address{Physics Department, State University of
New York at Buffalo, Buffalo, NY 14260, USA},
        A.~Juste\address{Fermi National Accelerator Laboratory,
Batavia, IL 60510, USA},
        L.~H.~Orr\address[MCSD]{Department of Physics and Astronomy, University
of Rochester, Rochester, NY 14627, USA}
        and
        D.~Rainwater\addressmark[MCSD]}
       
\runtitle{2-column format camera-ready paper in \LaTeX}
\runauthor{U.~Baur {\it et al.}}

\begin{document}

\begin{abstract}
We discuss possibilities to measure the $tt\gamma$ and $ttZ$ couplings
at hadron and lepton colliders. We also briefly describe how these
measurements can be used to constrain the parameter space of models of
new physics, in particular Little Higgs models.
\vspace{1pc}
\end{abstract}

\maketitle

\section{Introduction} 

Although the top quark was discovered almost ten years
ago~\cite{topcdf,topd0}, many of its properties are still only poorly
known~\cite{Chakraborty:2003iw}.  In particular, the couplings of the
top quark to the electroweak (EW) gauge bosons have not yet been
directly measured.  
Current data provide only weak constraints on the couplings of the top
quark with the EW gauge bosons, except for the $ttZ$ vector and axial
vector couplings which are rather tightly but indirectly constrained
by LEP data; and the right-handed $tbW$
coupling, which is severely bound by the observed $b\to s\gamma$
rate~\cite{Larios:1999au}.  

At an $e^+e^-$ linear collider with $\sqrt{s}=500$~GeV and an
integrated luminosity of $100-200$~fb$^{-1}$ one can hope to measure
the $ttV$ ($V=\gamma,\,Z$) couplings in top pair production with a few-percent
precision~\cite{Abe:2001nq}.  However, the process
$e^+e^-\to\gamma^*/Z^*\to t\bar{t}$ is sensitive to both $tt\gamma$
and $ttZ$ couplings and significant cancellations between the various
couplings can occur.  At hadron colliders, $t\bar{t}$ production is so
dominated by the QCD processes $q\bar{q}\to g^*\to t\bar{t}$ and
$gg\to t\bar{t}$ that a measurement of the $tt\gamma$ and $ttZ$
couplings via $q\bar{q}\to\gamma^*/Z^*\to t\bar{t}$ is hopeless.
Instead, the $ttV$ couplings can be measured in QCD $t\bar{t}\gamma$
production, radiative top quark decays in $t\bar{t}$ events
($t\bar{t}\to\gamma W^+W^- b\bar{b}$), and QCD $t\bar{t}Z$
production~\cite{Baur:2004uw}. $t\bar{t}\gamma$ production and radiative
top quark decays 
are sensitive only to the $tt\gamma$ couplings, whereas $t\bar{t}Z$
production gives information only on the structure of the $ttZ$
vertex.  This obviates having to disentangle potential cancellations
between the different couplings.

Here we briefly discuss the measurement of the $ttV$
couplings at the LHC and compare the expected sensitivities with the
bounds which one hopes to achieve at an $e^+e^-$ linear collider. 

\section{General \boldmath{$ttV$} Couplings}

The most general Lorentz-invariant vertex function describing the
interaction of a neutral vector boson $V$ with two top quarks can be
written in terms of ten form factors~\cite{Hollik:1998vz}, which are
functions of the kinematic invariants.  In the low energy limit,
these correspond to couplings which multiply dimension-four or -five 
operators in an effective Lagrangian, and may be complex.  If $V$ is 
on-shell, or if $V$ couples to effectively massless fermions, the 
number of independent form factors is reduced to eight.  If, in 
addition, both top quarks are on-shell, the number is further reduced 
to four.  In this case, the $ttV$ vertex can be written in the form
\begin{eqnarray}
& & \!\!\!\!\!\!\!\!\!\Gamma_\mu^{ttV}(k^2,\,q,\,\bar{q}) =  -ie \biggl\{
  \gamma_\mu \, \left( F_{1V}^V(k^2) + \gamma_5F_{1A}^V(k^2) \right)
\nonumber \\
& & \!\!\!\!\!\!\!\!\!\!+ 
 \frac{\sigma_{\mu\nu}}{2m_t}~(q+\bar{q})^\nu  
   \left( iF_{2V}^V(k^2) + \gamma_5F_{2A}^V(k^2) \right) 
\biggr\} \, , \label{eq:anomvertex}
\end{eqnarray}
where $e$ is the proton charge, 
$m_t$ is the top quark mass, $q~(\bar{q})$ is the outgoing top
(anti-top) quark four-momentum, and $k^2=(q+\bar{q})^2$.  The terms
$F_{1V}^V(0)$ and $F_{1A}^V(0)$ in the low energy limit are the $ttV$ 
vector and axial vector form factors.  The coefficients 
$F_{2V}^\gamma(0)$ and $F_{2A}^\gamma(0)$ are related to the magnetic 
and ($CP$-violating) electric dipole form factors.

\section{\boldmath{${t\bar{t}\gamma}$} Production at the LHC}

The most promising channel for measuring the $tt\gamma$ couplings at the
LHC is $pp\to \gamma\ell\nu_\ell b\bar{b}jj$, which receives
contributions from $t\bar t\gamma$ production and ordinary $t\bar t$
production where one of the top quarks decays radiatively ($t\to
Wb\gamma$). In order to reduce the background, it is advantageous to
require that both $b$-quarks are tagged. We assume a combined efficiency
of $\epsilon_b^2=40\%$ for tagging both $b$-quarks.

The non-resonant $pp \to W(\to\ell\nu)\gamma b\bar{b}jj$ background and
the single-top backgrounds, $(t\bar{b}\gamma + \bar{t}b\gamma)+X$, can
be suppressed by imposing invariant and transverse mass cuts which
require that the event is consistent either 
with $t\bar{t}\gamma$ production, or with $t\bar{t}$ production with
radiative top decay~\cite{Baur:2004uw}. Imposing a large
separation cut of $\Delta R(\gamma,b)>1$ reduces photon
radiation from the $b$ quarks.  Photon emission from $W$ decay
products can essentially be eliminated by requiring that
$m(jj\gamma) > 90~{\rm GeV}$ and $
m_T(\ell\gamma;p\llap/_T) > 90~{\rm GeV,}$
where $m(jj\gamma)$ is the invariant mass of the $jj\gamma$ system, and
$m_T(\ell\gamma;p\llap/_T)$ is the $\ell\gamma p\llap/_T$ cluster
transverse mass, which peaks sharply at $m_W$. 
After imposing the cuts described above, the irreducible backgrounds are
one to two orders of magnitude smaller than the signal.

The potentially most dangerous reducible background is $t\bar{t}j$ 
production where one of the jets in the final state fakes a photon.
\begin{figure*}
\begin{center}
\begin{tabular}{cc}
\includegraphics[width=7.7cm]{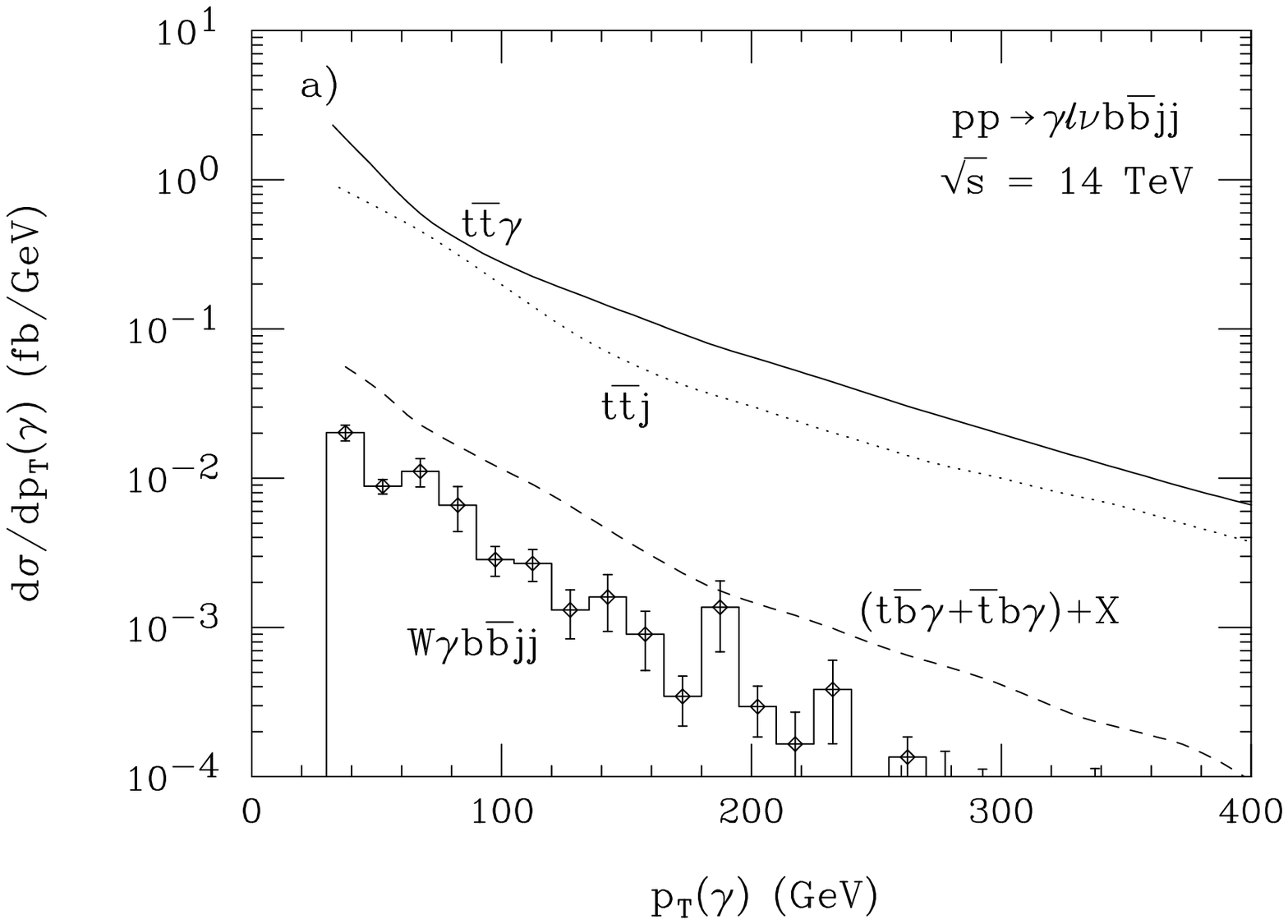} &
\includegraphics[width=7.7cm]{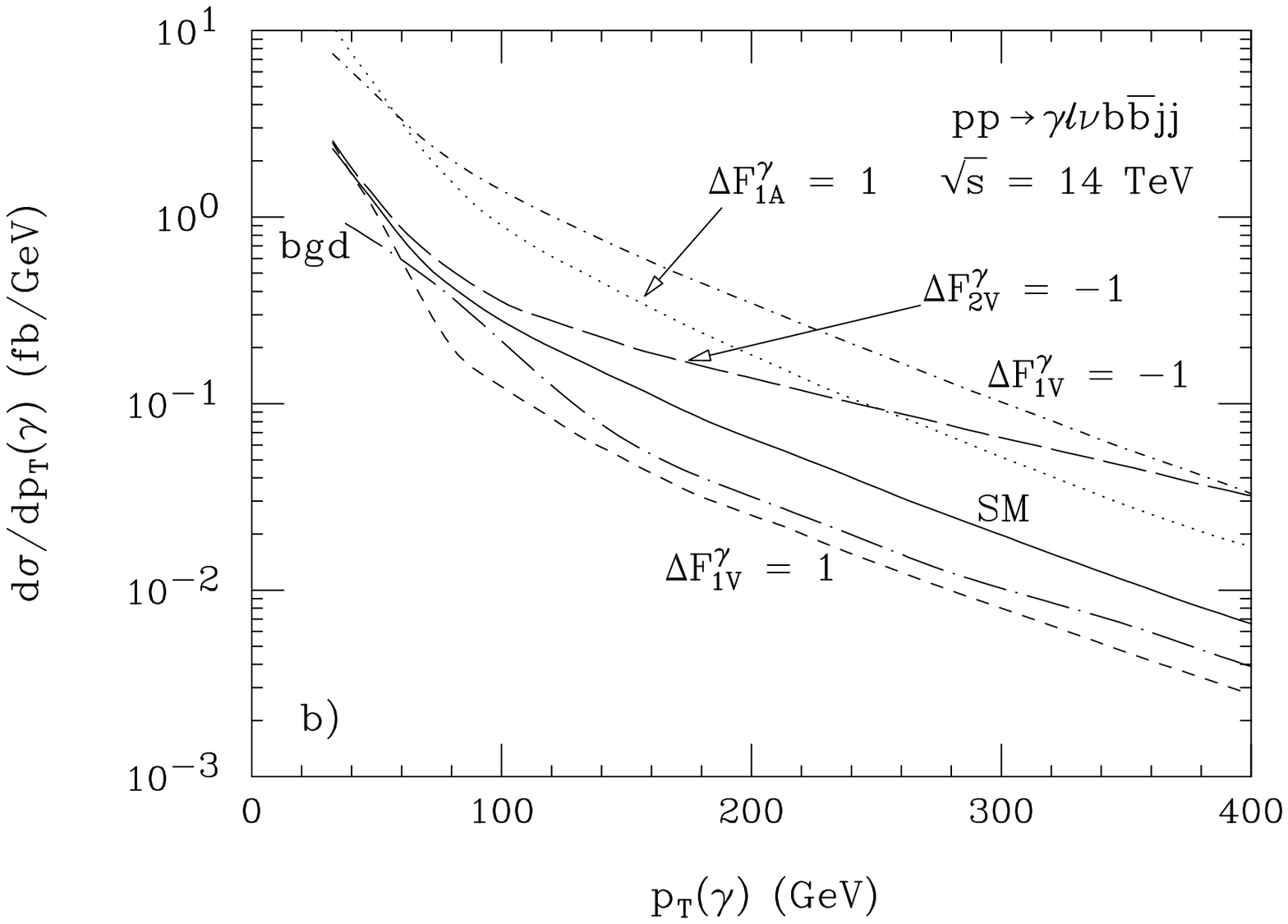}
\end{tabular}
\end{center}
\vskip -1.cm
\caption{\label{fig:fig1}{The differential cross sections as a function
of the photon  transverse momentum for $\gamma\ell\nu_\ell b\bar{b}jj$
production at the LHC. Part a) shows the SM signal and the various
contributions to the background. Part b) shows the SM signal and
background, and the signal for various anomalous $tt\gamma$ couplings.}}
\end{figure*}
In Fig.~\ref{fig:fig1}a we show the photon transverse momentum
distributions of the $t\bar{t}\gamma$ signal and the backgrounds
discussed above. The $t\bar{t}j$ background is seen to
be a factor~2 to~3 smaller than the $t\bar{t}\gamma$ signal for the
jet -- photon misidentification probability
($P_{j\to\gamma}=1/1600$~\cite{atlas_tdr}) used.  

The photon transverse momentum distributions in the SM and for various
anomalous $tt\gamma$ couplings, 
together with the $p_T(\gamma)$ distribution of the
background, are shown in
Fig.~\ref{fig:fig1}b. Only one coupling at a time is allowed to deviate
from its SM prediction. 

\section{\boldmath{${t\bar{t}Z}$} Production at the LHC}

The process $pp \to t\bar{t}Z$ leads
to either ${\ell'}^+{\ell'}^-\ell\nu b\bar{b}jj$ or
${\ell'}^+{\ell'}^- b\bar{b}+4j$ final states if the $Z$-boson
decays leptonically and one or both of the $W$ bosons decay
hadronically. If the $Z$ boson decays into neutrinos and both $W$ bosons
decay hadronically, the final state consists of
$p\llap/_Tb\bar{b}+4j$. Since there is essentially no phase space for
$t\to WZb$ decays ($BR(t\to WZb)\approx 3\cdot
10^{-6}$~\cite{Mahlon:1994us}), these final states
arise only from $t\bar tZ$ production. 

In order to identify leptons, $b$ quarks, light jets and the missing
transverse momentum in dilepton and trilepton events, the same
cuts as for $t\bar t\gamma$ production are imposed. One also requires that
there is a same-flavor, opposite-sign lepton pair with 
invariant mass near the $Z$ resonance, $m_Z - 10~{\rm GeV} < m(\ell\ell)
< m_Z + 10~{\rm GeV}$. 

The main backgrounds contributing to the trilepton final state are
singly-resonant $(t\bar{b}Z+\bar{t}bZ)+X$ ($t\bar{b}Zjj$,
$\bar{t}bZjj$, $t\bar{b}Z\ell\nu$ and $\bar{t}bZ\ell\nu$) and
non-resonant $WZb\bar{b}jj$ production. In the dilepton case, the main
background arises from 
$Zb\bar{b}+4j$ production.  To adequately suppress it, one
additionally requires that events have at least one combination of jets
and $b$ quarks which is consistent with the $b\bar b+4j$ system
originating from a $t\bar t$ system. Once these cuts have been imposed,
the $Zb\bar{b}+4j$ background is important only for $p_T(Z)<100$~GeV. 

\begin{figure*}
\begin{center}
\begin{tabular}{cc}
\includegraphics[width=7.7cm]{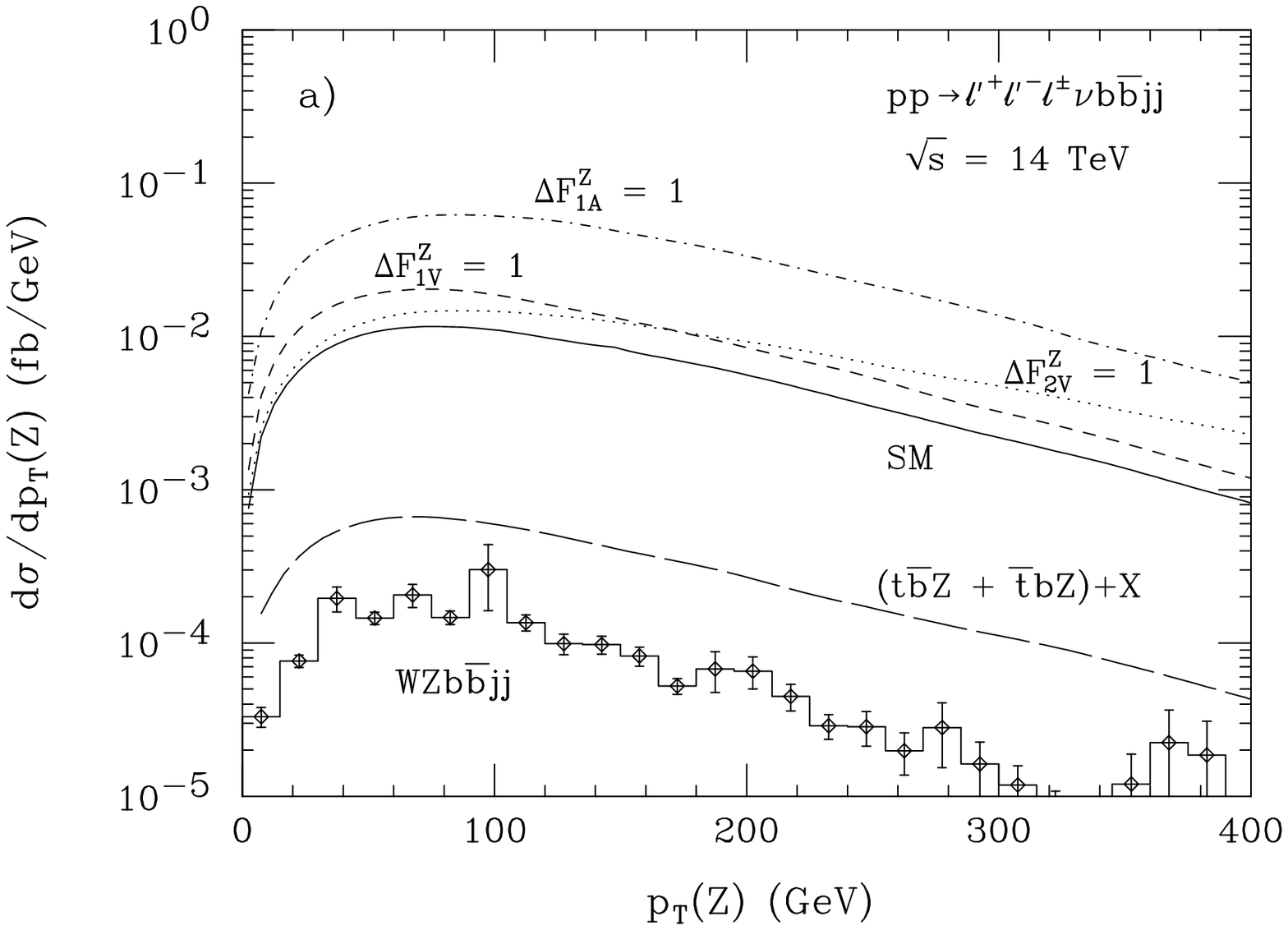} &
\includegraphics[width=7.7cm]{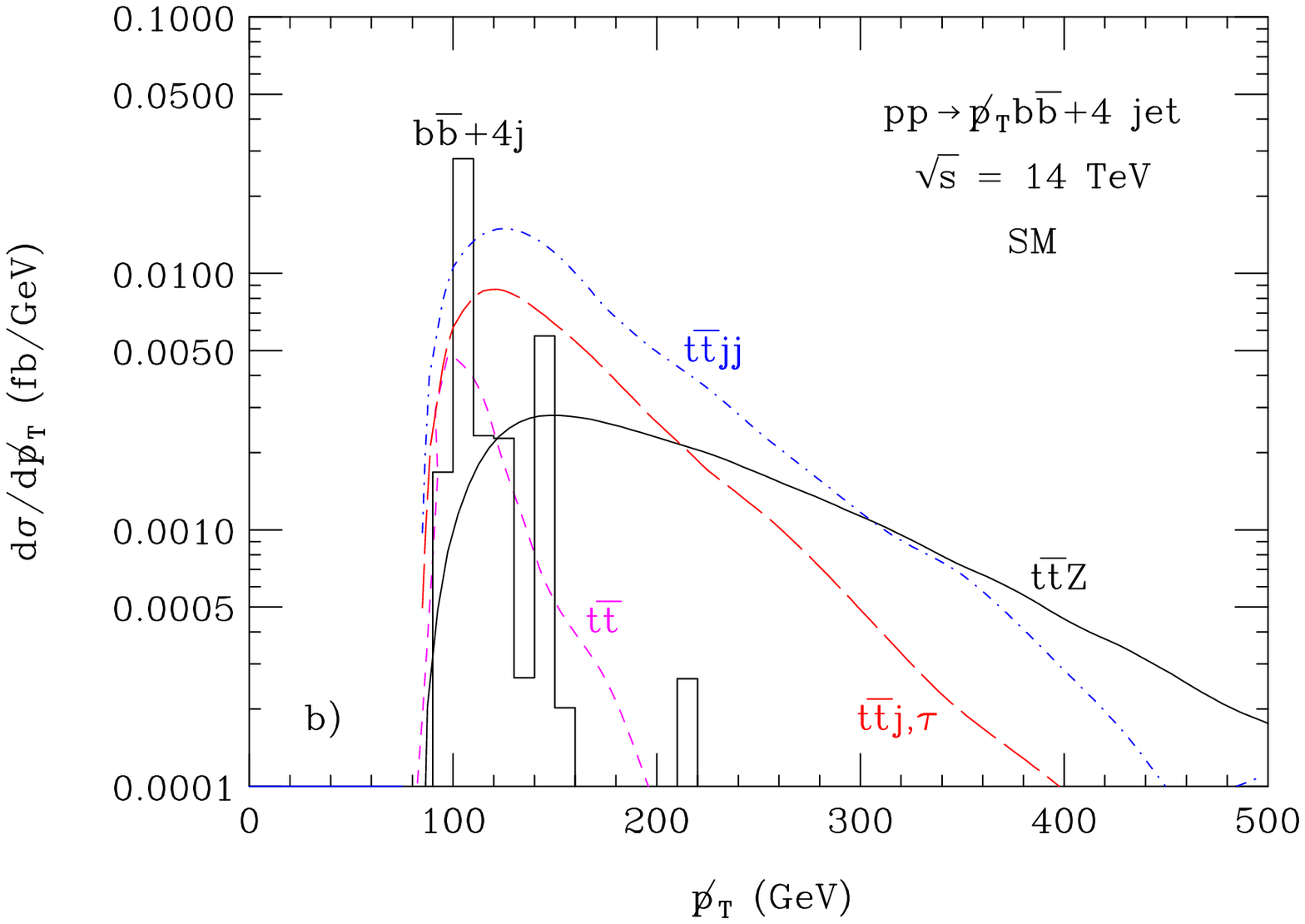}
\end{tabular}
\end{center}
\vskip -1.cm
\caption{\label{fig:fig2}{a) The differential cross sections at the LHC
as a function 
of $p_T(Z)$ for ${\ell'}^+{\ell'}^-\ell\nu b\bar{b}jj$ final states.  
Shown are the SM predictions for $t\bar{t}Z$ production, for several
non-standard $ttZ$ couplings, and for various
backgrounds.  Only one coupling at a time is allowed to deviate from 
its SM value. 
b) The differential cross sections as a function of the missing
transverse momentum for $p\llap/_Tb\bar b+4$j production at the
LHC.  Shown are the SM predictions for
$t\bar{t}Z$ production and for various backgrounds. }}
\end{figure*}
The $Z$ boson transverse momentum distribution for the trilepton final
state is shown in
Fig.~\ref{fig:fig2}a for the SM signal and backgrounds, as well as for 
the signal with several non-standard $ttZ$ couplings. Only one coupling
at a time is allowed to deviate from its SM 
prediction.  The backgrounds are each more than one order of magnitude
smaller than the SM signal. Note that varying
$F^Z_{1V,A}$ leads mostly to a cross section 
normalization change, hardly affecting the shape of the $p_T(Z)$
distribution.  

For the $p\llap/_Tb\bar{b}+4j$~\cite{new} final state at least 3~jets 
with $p_T>50$~GeV and $p\llap/_T>5~{\rm GeV}^{1/2}\sqrt{\sum p_T}$ are
required. The 
largest backgrounds for this final state come from $t\bar t$ and $b\bar
b+4j$ production where one or several jets are badly mismeasured, from
$pp\to t\bar tjj$ with $t\bar 
t\to\ell^\pm\nu_\ell b\bar bjj$ and the charged lepton
being missed, and from $t\bar tj$ production, where one top decays
hadronically, $t\to Wb\to bjj$, and the other via $t\to Wb\to\tau\nu_\tau b$
with the $\tau$-lepton decaying hadronically, $\tau\to h\nu_\tau$. 

In Fig.~\ref{fig:fig2}b we show the missing transverse momentum
distributions of the SM $t\bar tZ\to p\llap/_Tb\bar{b}+4j$ signal (solid
curve) and various 
backgrounds. The most important backgrounds are $t\bar tjj$
and $t\bar tj$ production. However, the missing transverse momentum
distribution from these processes falls considerably faster than that of
the signal, and for $p\llap/_T>300$~GeV, the SM signal dominates. 

\section{Sensitivity Bounds for $ttV$ Couplings: LHC and ILC}

The shape and normalization changes of the photon or $Z$-boson
transverse momentum distribution can be used to derive
sensitivity bounds on the anomalous $tt\gamma$ and $ttZ$
couplings. For $t\bar{t}Z$ production with $Z\to{\ell}^+{\ell}^-$, the
distribution of the ${\ell}^+{\ell}^-$ opening angle in the transverse
plane, $\Delta\Phi({\ell}{\ell})$,
provides additional information~\cite{Baur:2004uw}. In the
following we assume a normalization uncertainty of the SM cross
section of $\Delta{\cal N}=30\%$. 

Even for a modest integrated luminosity of 30~fb$^{-1}$, it will be 
possible to measure the
$tt\gamma$ vector and axial vector couplings, and the dipole form
factors, with a precision of typically $20\%$ and $35\%$,
respectively. For 300~fb$^{-1}$, the limits improve to $4-7\%$ for
$F^\gamma_{1V,A}$ and to about $20\%$ for $F^\gamma_{2V,A}$. 

To extract bounds on the $ttZ$ couplings, we perform a simultaneous
fit to the $p_T(Z)$ and the $\Delta\Phi({\ell'}{\ell'})$
distributions for the trilepton and dilepton final states, and to the
$p\llap/_T$ distribution for the $p\llap/_Tb\bar b+4j$ final state. 
For an integrated luminosity of 300~fb$^{-1}$, it will be possible to
measure the $ttZ$ axial vector coupling with a precision of $10-12\%$,
and $F^Z_{2V,A}$ with a precision of $40\%$.  At the SLHC, assuming an
integrated luminosity of 3000~fb$^{-1}$, these 
bounds can be improved by factors of about~1.6 ($F^Z_{2V,A}$)
and~3 ($F^Z_{1A}$).  The bounds which can be achieved for
$F^Z_{1V}$ are much weaker than those projected for $F^Z_{1A}$.  As
mentioned in Sec.~4, the $p_T(Z)$ distributions for the
SM and for $F^Z_{1V,A}=-F^{Z,SM}_{1V,A}$ are almost degenerate.
This is also the case for the $\Delta\Phi({\ell'}{\ell'})$
distribution.  In a fit to these two distributions, therefore, an area
centered at $\Delta F^Z_{1V,A}=-2F^{Z,SM}_{1V,A}$ remains which cannot
be excluded, even at the SLHC.  For $F^Z_{1V}$, the two regions merge,
resulting in rather poor limits.  

It is instructive to compare the bounds for anomalous $ttV$ couplings
achievable at the LHC with those projected for the ILC. The most
complete study of $t\bar{t}$ production at the ILC for general $ttV$
($V=\gamma,\,Z$) couplings so far is 
that of Ref.~\cite{Abe:2001nq}.  Note that only one coupling at a time
is allowed to deviate from its SM value in
Ref.~\cite{Abe:2001nq}. Comparing the projected LHC and ILC sensitivity
bounds, one finds~\cite{snowmass}
that, even if the SLHC operates first, and the
$p\llap/_Tb\bar b+4j$ final state is taken into account, a linear
collider will still be able to significantly improve the $ttZ$ anomalous
coupling 
limits, with the possible exception of $\widetilde F^Z_{1A}$. The ILC
will also be able to considerably strengthen the bounds on $\widetilde
F^\gamma_{1A}$ and $\widetilde F^\gamma_{2A}$. It should
be noted, however, that this 
picture could change once correlations between different non-standard
$ttZ$ couplings, and between $tt\gamma$ and $ttZ$ couplings, are taken
into account. Unfortunately, so far no realistic studies for $e^+e^-\to
t\bar t$ which include these correlations have been performed.

\section{Model Implications}

Many models of new physics predict anomalous $ttZ$ couplings. Examples
are top-seesaw models~\cite{Dobrescu:1997nm} and Little Higgs 
models~\cite{LH}, which predict an up-type 
quark singlet $T$ which mixes with the top quark. This changes
coupling of the left-handed top quark to the $Z$-boson:
\begin{equation}
\Delta F^Z_{1V}=-\Delta F^Z_{1A}=F^{Z,SM}_{1A}\sin^2\theta_L\,
\end{equation}
where $\theta_L$ is the $T-t$ mixing angle.

It is straightforward to derive bounds for $\sin^2\theta_L$ from the
general limits on $F^Z_{1V,A}$ outlined in Sec.~5. With 300~fb$^{-1}$,
$\sin^2\theta_L$ can be restricted to
\begin{equation}
\sin^2\theta_L< 0.084~(0.16) \qquad {\rm at~68.3\%~(95\%)~CL}.
\end{equation}
At the SLHC it will be possible to improve these bounds by about a
factor~2.8. 

In the $SU(5)/SO(5)$ Littlest Higgs model with
T-parity~\cite{Hubisz:2005tx}, $\sin^2\theta_L$ is related to the mass
of the heavy top quark
partner, $T$, the $tTh$ 
coupling ($h$ is the Higgs boson), $\lambda_T$, and the SM Higgs vacuum
expectation value, $v\approx 246$~GeV, by
\begin{equation}
\sin^2\theta_L={\lambda_T^2v^2\over 2m^2_T}.
\end{equation}
In this model, the bounds on $\sin^2\theta_L$ can be converted into
limits on $m_T/\lambda_T$. For 300~fb$^{-1}$ one finds~\cite{BPP}
\begin{equation}
{m_T\over\lambda_T}>600~(430)~{\rm GeV} \qquad {\rm at~68.3\%~(95\%)~CL}.
\end{equation}
Since the LHC should be able to discover a $T$ quark with a mass of
$m_T\leq 2$~TeV~\cite{Azuelos:2004dm}, a measurement of $F^Z_{1A}$ can
provide valuable information on $\lambda_T$. 

\section{Conclusions}

The LHC will be able to perform first tests of the $ttV$ couplings. 
Already with an
integrated luminosity of 30~fb$^{-1}$, one can probe the $tt\gamma$ couplings
with a precision of about $10-35\%$ per experiment.  With higher
integrated luminosities one will be able to reach the few percent
region. The LHC will also be able to probe the $ttZ$ couplings, albeit
not with the same level of precision. However, 
a measurement of $F^Z_{1A}$ will constrain the parameter space
of models with an extra up-type singlet quark, such as Little Higgs
models. The ILC will
be able to further improve our knowledge of the $ttV$ couplings, in
particular in the $ttZ$ case.

\section*{Acknowledgments}

This material is based upon work supported by the Department of
Energy under Award Number DE-FG02-91ER40685.
This research was also supported in
part by the National Science Foundation under grant No.~PHY-0456681.


\end{document}